\documentclass[aps,pre,twocolumn,showpacs,preprintnumbers,amsmath,amssymb,nofootinbib,floatfix]{revtex4-1} 
\usepackage{graphicx,color}  
\usepackage{subfig}     
\usepackage{amssymb}   
\usepackage{amsmath}   
\begin{document}
\newcommand{\kbt}{k_{\mathrm{B}}T}


\title{Interfacial Phenomena of Solvent-diluted Block Copolymers}
\author{Shai Cohen}
\author{David Andelman} \affiliation{Raymond and Beverly Sackler School
of Physics and Astronomy, Tel Aviv University, Ramat Aviv, Tel Aviv 69978, Israel}

\vskip 0.25cm

\date{submitted: Dec 16, 2013} 

\begin{abstract}

 A phenomenological mean-field theory is used to investigate the properties of solvent-diluted
di-block copolymers (BCP), in which the two BCP components (A and B) form a variety of phases that are
diluted by a solvent (S). Using this approach, we model mixtures of di-block copolymers and
a solvent and obtained the corresponding critical behavior.
In the low solvent limit, we find how the
critical point depends on the solvent density. Due to the non-linear nature of the coupling between the A/B and BCP/solvent concentrations, the A/B modulation induces modulations in the polymer-solvent relative concentration with a double wavenumber. The free boundary separating the polymer-rich
phase from the solvent-rich one is studied in two situations. First, we show how the presence of a chemically patterned substrate leads to deformations of the BCP film/solvent interface, creation of terraces in lamellar BCP film and even formation of multi-domain droplets as induced by the patterned substrate. Our results are in agreement with previous self-consistent field theory calculations. Second, we compare the surface tension between parallel lamellae coexisting with a solvent phase with that of a perpendicular one, and show that the surface tension has  a non-monotonic dependence on temperature. The anisotropic surface tension  can lead to deformation of spherical BCP droplets into lens-shaped ones, together with re-orientation of the lamellae inside the droplet during the polymer/solvent phase separation process in agreement with experiment.
\end{abstract}

\maketitle

\section{Introduction}

A wide variety of chemical and physical systems exhibiting patterns and textures can be characterized by
spatial modulations in thermodynamical equilibrium. Some of the most common morphologies are
stripes and circular droplets in two dimensions (2d), as well as sheets, tubes and
spherical droplets in three dimensions (3d). These systems are very diverse and include
type I superconductor films, ferromagnetic films, block copolymers (BCP),
and even lipid bio-membranes~\cite{Seul_DA_96,DA_RER1,DA_RER2}. In each case the physical origin of the order
parameter and the pattern characteristic differ with length scales or periodicity that
vary from mesoscopic scales of tens of nanometers in biological
membranes~\cite{membranes}, to centimeters in ferrofluids~\cite{DA_RER1,DA_RER2}. The fact
that such a vast variety of physical, chemical, and biological systems display morphological
similarities, irrespective of the details of microscopic structure and interactions, is striking and
can be explained by a generic mechanism of competing interactions
\cite{Seul_DA_96,DA_RER1,DA_RER2}.

A Ginzburg-Landau (GL) expansion~\cite{GL_tricritical}
applicable to modulated phases was introduced~\cite{Brazovskii} by adding  to the free energy a term preferring a specific
wavelength $q_0$. This GL-like approach works rather well close to the critical point (weak segregation limit), and its advantage lies in its simplicity and analytical predictions.
The added term in the GL free-energy expansion can be written as:
\begin{equation}
\frac{1}{2}\int\bigg[(\nabla^2+q^2_0)\phi\bigg]^2d^2r,
\label{Brazovskii.e.q.}
\end{equation}
This positive-definite form is known as the Brazovskii form
\cite{Brazovskii,Leibler,Fluctuation_BCP,Fredrickson_Binder,Landau-Brazovskii,Tsori_01,Thiele13}, and has a minimum at $q_0$. Close to the critical point (the weak segregation limit),
also called the Order-Disorder Temperature (ODT) in BCP systems,
it has been shown by Brazovskii~\cite{Brazovskii} and Leibler~\cite{Leibler}
that the amplitude of the most dominant $q$-mode, $q_0$, grows much faster than other $q$-modes.
This free-energy form and similar ones have been used extensively in the past to
calculate phase diagrams and grain boundaries of modulated phases~\cite{Netz_PRL_97,Phase-Transition,Tsori_chevron00},
as well as to study the effects of
chemically patterned surfaces on such phases~\cite{Fredrickson_87,Tsori_01a,Tsori_05,Tsori_03,para_perp,Matsen97,Geisinger99}.
In particular, such coarse-grained models compare well
with experimental phase diagrams~\cite{BCP_phase_diag95}
and grain boundaries studies of BCP~\cite{Hashimoto93,Gido93,Gido94}. Other and
more accurate numerical schemes exist and rely on self-consistent field theories (SCFT), as
well as Monte Carlo simulations. For a recent review of these complementary techniques see, e.g., ref~\cite{Detcheverry_softmatter09}.

In the present study we consider an extension of modulated phases that are diluted by a solvent.
Although our phenomenological approach can apply to any modulated phase, we apply it explicitly to symmetric lamellar phases of BCP diluted by a solvent in a variety of solvent conditions~\cite{X_Man_PRE12}.
For example, in bad solvent conditions the phase separation between a BCP-rich phase and a solvent-rich phase
allows us to explore the free interface between a BCP film and bad solvent (vapor). Our study is relevant to a large number of experimental situations where a BCP film is spin casted on a solid substrate~\cite{Coulon89,Knoll02,Knoll04,Stoykovich05,Ruiz08,Bang09,Segalman01,Stein07,Li04,Voet11,Thebault12}. As the solvent evaporates, the polymer/air free interface can deform, and the free interface  self-adjust its shape in order to minimize the total free energy. Another focus of our study is to consider the shape and orientation of multi-lamellar BCP domains, during phase separation between the polymer and a bad solvent.

The outline of our paper is as follows. In the next section we present our model, which is a generalization of phenomenological free-energy expansions used to model pure di-BCP systems. The free energy includes additional terms that depend on the solvent (S) and its coupling with the A/B relative composition. In section III we present the bulk properties of the solvent-diluted BCP system, including an explanation of the period-doubling phenomena found for the polymer density and an analysis of the critical point in the low-solvent limit. In section IV, we present results for  grain boundaries between lamellar domains of different orientations, and show how a chemically patterned substrate influences deformations of a flat free interface and formation of BCP droplets. We then proceed by calculating the temperature dependence of the lamellar-solvent surface tension for parallel and perpendicular lamellar orientations. We also show how such an anisotropic surface tension affects the shape of circular lamellar drops of BCP during the overall solvent/polymer phase separation process and compare with experiments.

\section{The Free-Energy of Mixing}

In this study we employ a phenomenological free-energy of mixing~\cite{Rubinstein,Doi}
of a ternary mixture composed of three
components: $\textrm{A}$,\;$\textrm{B}$ and $\textrm{S}$. The $\textrm{A}$ and $\textrm{B}$
components are the two components of a di-BCP~\cite{Netz_PRL_97} and have volume fractions $\phi_\textrm{A}$
and $\phi_\textrm{B}$, while $\textrm{S}$ is a solvent with volume fraction
$\phi_\textrm{S}$.
All three volume fractions satisfy $0\leq\phi_\textit{i}\leq1$,\;
$\textit{i}=\textrm{A},\textrm{B},\textrm{S}$.

Using the incompressibility condition
\begin{equation}
\phi_\textrm{A}+\phi_\textrm{B}+\phi_\textrm{S}=1 \,,
\label{eq_incompressibility}
\end{equation}
it is more convenient to consider the two following order-parameters:
\begin{eqnarray}
\rho&=&\phi_\textrm{A}+\phi_\textrm{B} \, , \nonumber\\
\phi&=&\phi_\textrm{A}-\phi_\textrm{B} \, ,
\end{eqnarray}
where $0\le \rho\le 1$ is the solute (BCP) volume fraction and $-1\le \phi \le 1$
is the relative $\textrm{A/B}$ concentration of the two blocks, satisfying $|\phi|\le \rho$.

We work in the grand-canonical ensemble where the  Gibbs free-energy density (per unit volume)
$g$ is defined as
\begin{equation}
g=\varepsilon-Ts-\sum_{i=\textrm{A,B,S}}\mu_i\phi_i \, ,
\end{equation}
with $T$ being the temperature, $\varepsilon$ the enthalpy, $s=-k_B\sum_{i}\phi_i\ln(\phi_i)$  the ideal entropy of mixing, and $\mu_i$  the chemical potential of
the $i$-th species. Writing  $\varepsilon$ in terms of
all two-body interactions between the three components, and expressing the free-energy $g$ in terms of the $\rho$ and $\phi$ densities, yields
\begin{eqnarray}
\frac{g(\phi,\rho)}{\kbt}&=&-\frac{\tau}{2}\phi^2+\frac{\chi}{2}\rho(1-\rho)+v_{\phi\rho}\phi(1-\rho)-\mu_{\phi}\phi -\mu_{\rho}\rho\nonumber\\
& & +~\frac{\rho+\phi}{2}\ln(\rho+\phi)
+\frac{\rho-\phi}{2}\ln(\rho-\phi)\nonumber \\
& &+~(1-\rho)\ln(1-\rho)
+\frac{H}{2}\bigg[(\nabla^2+q^2_0)\phi\bigg]^2\nonumber \\ & & +~K(\nabla\rho)^2 \, ,
\label{eq_full-free-energy}
\end{eqnarray}
where $\tau$ is the interaction parameter between the \textrm{A} and \textrm{B} monomers of the BCP,
$\chi$ is the solvent-polymer interaction parameter, and $v_{\phi\rho}$ is the parameter denoting any asymmetry in the interaction
between the solvent and the A and B components. For simplicity, throughout this paper we  set $v_{\phi\rho}=0$, modeling
only symmetric interactions of the A and B components with a neutral S solvent. Clearly that in this case all our results will
obey the symmetry $\phi \leftrightarrow -\phi$.

The next two terms are the chemical potential ones, where $\mu_\phi$ couples linearly to $\phi$ and
$\mu_\rho$ to the polymer volume fraction $\rho$, and the three logarithmic terms
originate from the ideal entropy of mixing.
So far the terms of the free energy, $g$,  describe any three-component mixture of A, B and S within the ideal solution (mean-field) theory.
In order to model the A/B mixture as a di-BCP, we add to eq~\ref{eq_full-free-energy}
the $H$-term as introduced in eq~\ref{Brazovskii.e.q.},
where $H$ is the modulation coefficient and $q_0$ is the most dominant wavevector~\cite{Netz_PRL_97,Phase-Transition,Brazovskii,Leibler,Fluctuation_BCP,
Fredrickson_Binder,Landau-Brazovskii,Tsori_01}. As we are interested in studying interfacial phenomena between polymer-rich and solvent-rich phases,
we also included a gradient squared term in the polymer density $\rho$ to account for
the cost of density fluctuations, where $K>0$ is a measure of the interface ``stiffness"
\cite{Safran}.

Note that for the symmetric $v_{\phi\rho}=0$ case, the
only coupling terms in eq~\ref{eq_full-free-energy} between the two order parameters,
$\phi$ and $\rho$, comes from the entropy,
and are even in $\phi$. There is no bilinear term $\phi\rho$ and
the lowest-order coupling term is $\phi^2\rho$. The free energy
can be reduced to two simple limits.
(i) For $\rho=1$, the system reduces to the pure A/B BCP. Here $\phi_S=0$ and
\begin{eqnarray}
\frac{g}{\kbt}&=&-\frac{\tau}{2}\phi^2+\frac{1+\phi}{2}\ln(1+\phi)
+\frac{1-\phi}{2}\ln(1-\phi)\nonumber \\
& & +~\frac{H}{2}\bigg[(\nabla^2+q^2_0)\phi\bigg]^2-\mu_{\phi}\phi \, .
\label{eq_rhoEq1}
\end{eqnarray}
This case was mentioned in section I and has been studied
extensively in the past for pure BCP systems~\cite{Brazovskii,Leibler}. The BCP phase has a critical point (ODT) at $\tau_c=1$ and for $\tau>\tau_c$, only
the disordered phase is stable.

(ii) Another simple limit is obtained by setting $\phi_0=0$ (or $\phi_\textrm{A}=\phi_\textrm{B}$), yielding
\begin{eqnarray}
\frac{g}{\kbt}&=&K(\nabla\rho)^2+\frac{\chi}{2}\rho(1-\rho)+\rho\ln{\rho}\nonumber\\
& & +~(1-\rho)\ln(1-\rho)-\mu_{\rho}\rho \, .
\end{eqnarray}
This is the free energy of a solvent/solute binary mixture, where $\rho$ is the solute volume fraction~\cite{Rubinstein}. In the bulk, $\rho=const$, and one gets a demixing curve between two macroscopically separated phases. The demixing curve terminates
at a critical point located at $\chi_\textrm{c}=4$
and $\rho_\textrm{c}=0.5$.

\section{Bulk Properties}

\subsection{The Low Solvent Limit and Criticality}

\begin{figure}[!ht]
  \subfloat{\includegraphics[width=0.35\textwidth] {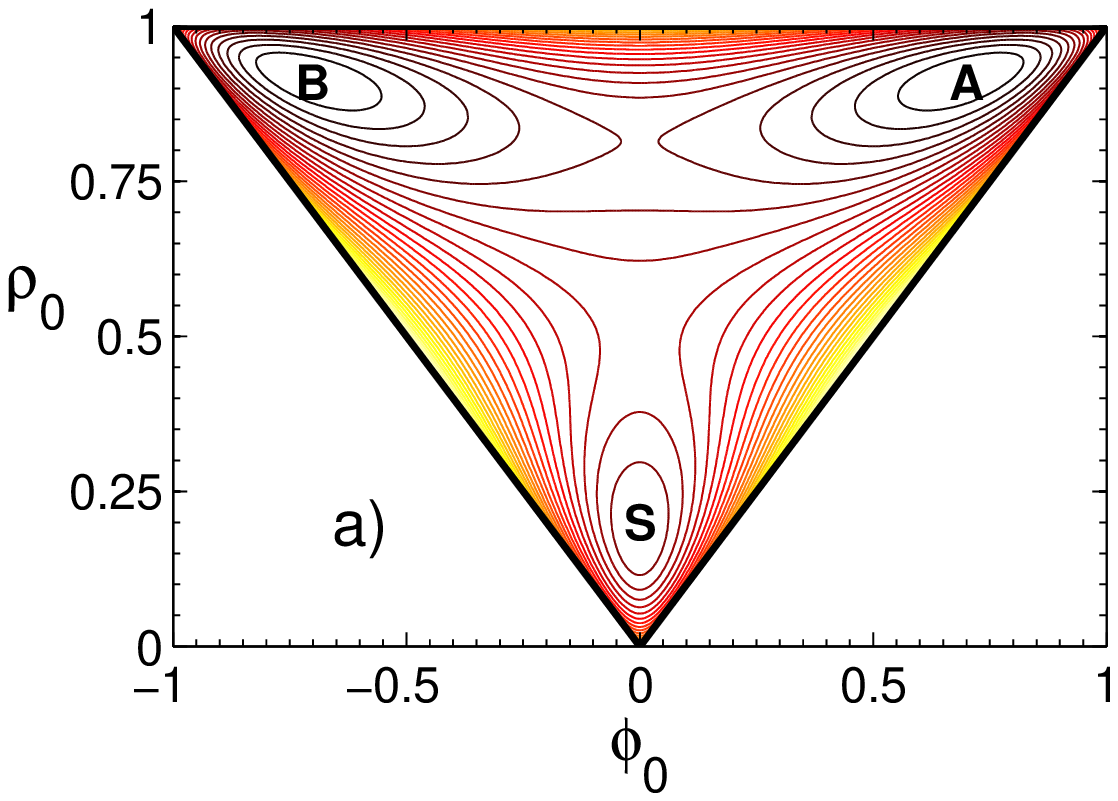}} 
  \quad
  \subfloat{\includegraphics[width=0.35\textwidth] {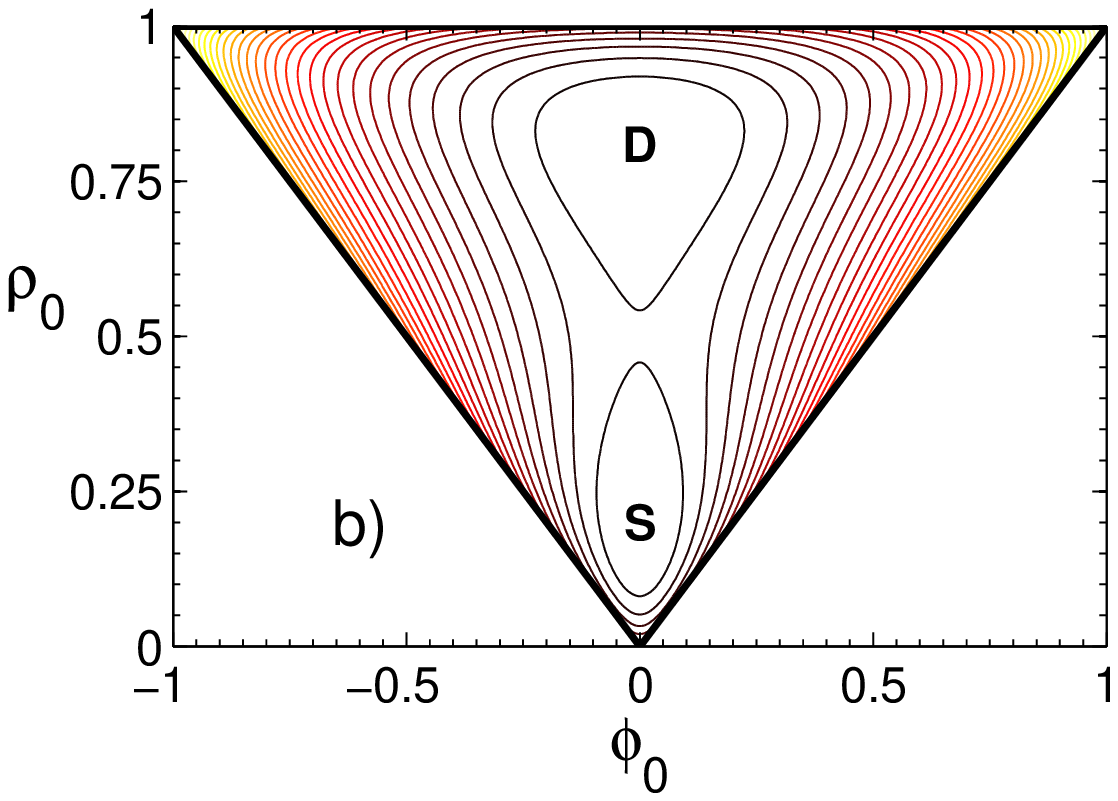}} 
  \caption{\textsf{(colored online) The free-energy landscape in the ($\phi_0$,$\rho_0$) plane.
  The contour plot is obtained from the ternary-mixture
free energy, eq~\ref{eq_full-free-energy}, disregarding the spatially varying terms, for $\chi=4.4$ and $\mu_{\rho}=\mu_{\phi}=0$. In (a) $\tau=1.45>\tau_c$ while in (b) $\tau=0.5<\tau_c$. In (a) the free-energy minima are denotes by A, B and S, and in (b) by D and S. }}
  \label{fig1}
\end{figure}

Since our free energy reduces to eq~\ref{eq_rhoEq1} for $\rho_0=1$ (no solvent),
a BCP phase for the pure A/B system exists as long as $\tau>\tau_{\textrm{c}}$,
where $\tau_c(\rho_0=1)=1$ is the critical point (ODT).
To gain some
insight of the behavior for the ternary system with $\rho_0<1$, we plot in figure~\ref{fig1} typical energy landscapes
corresponding to the free energy $g$ of eq~\ref{eq_full-free-energy} without the spatial-dependent
$H$ and $K$ terms.
In figure~\ref{fig1}(a), $\tau=1.45$ is above the critical point, $\tau_c=1$,
and we notice three local minima denoted by
$\textrm{A}$, $\textrm{B}$ and $\textrm{S}$. The $\textrm{A}$ and $\textrm{B}$ minima denote
two equivalent points for which the solute density is high ($\rho_0\approx 0.9$) and is
composed mainly of the $\textrm{A}$ component ($\phi_0\approx 0.7$) or the $\textrm{B}$ component ($\phi_0\approx -0.7$).
Point $\textrm{S}$, on the other hand, has low solute density ($\rho_0\simeq0.2$) and is
composed of an equal amount of the $\textrm{A}$ and $\textrm{B}$  $(\phi_0=0)$. In
addition, note also that the free energy is concave on the $\rho_0=1$ line.

Figure \ref{fig1}(a) should be compared with \ref{fig1}(b), for which $\tau=0.5 < \tau_c=1$. In this
case, only two minima exist and are denoted as $\textrm{S}$ (solvent rich as before) and $\textrm{D}$ (disorder solute-rich).
For the D minimum the concentration $\rho_0$ is high and equal amounts of $\textrm{A}$ and
$\textrm{B}$ component are present $(\phi_0=0)$. The free energy is convex on the $\rho_0=1$ line.
We conclude that modulations can only occur if the $\textrm{A}$ and $\textrm{B}$
minima exist so that the two component tend to phase separate, while the modulation term dictates the
length scale of spatial modulations by having the $\phi$ composition oscillates between these two states.
Looking at figure~\ref{fig1}(a) along the symmetric $\phi_0=0$ line, we further remark that the free energy
changes from being slightly concave at $\rho_0 \lesssim 1$ to highly convex close to $\rho_0=0$. This
change in convexity suggests that a lamellar phase in $\phi$ will only be energetically
favorable at $\rho_0\simeq1$, because in the highly convex region only one minimum exists. It
also suggests that the critical point $\tau_{\textrm{c}}$ will grow as $\rho_0$ decreases below $\rho_0=1$.

Next, we would like  to obtain
an analytical expression of $\tau_{\textrm{c}}(\rho)$, for $\rho\lesssim 1$, while recalling that without any solvent, $\tau_c(\rho{=}1)=1$. We
use the lamellar single-mode  approximation, and expand the free energy in powers of $\phi/\rho$. Because $|\phi|<\rho$,
$\phi/\rho$ may serve as a small expansion parameter around the ($\rho_0\simeq 1$, $\phi_0=0$) corner of the
phase diagram. Expanding the entropy in eq~\ref{eq_full-free-energy} to fourth order in $\phi/\rho$ results in:
\begin{eqnarray}
\frac{g}{\kbt}&\simeq& -\frac{\tau}{2}\phi^2+\frac{\chi}{2}\rho(1-\rho)+(1-\rho)\ln(1-\rho)+
\rho\ln\rho\nonumber \\
& &+\frac{\phi^2}{2\rho}+\frac{\phi^4}{12\rho^3}+\frac{H}{2}\bigg[(\nabla^2+q^2_0)\phi\bigg]^2\nonumber\\
& &+~K(\nabla\rho)^2 -\mu_{\rho}\rho-\mu_{\phi}\phi. \label{phi expansion}
\end{eqnarray}

Since the only source of modulations is the $\frac{H}{2}[(\nabla^2+q^2_0)\phi]^2$ term,
modulations in $\rho$ can be induced by modulations in $\phi$ through the coupling between $\phi$ and $\rho$ (that in our model
originates only from the entropy terms)
as will be explain in detail in section~III.B below. Thus, it is reasonable to assume that close to the critical point, where $\phi$  starts to
modulate, the modulations in $\rho$ are small and their effect on the $\phi$ modulations can be neglected (to be further justified below).

Dropping the constant terms in $\rho$ from eq~\ref{phi expansion} yields,
\begin{equation}
\frac{g}{\kbt}=-\frac{\tau}{2}\phi^2+\frac{\phi^2}{2\rho}+\frac{\phi^4}{12\rho^3}
+\frac{H}{2}\bigg[(\nabla^2+q^2_0)\phi\bigg]^2 \,.
\end{equation}
Assuming for simplicity 1d lamellar modulations that are symmetric around $\phi_0=0$,
we use for $\phi$ the single-mode  ansatz $\phi(x)=\phi_0+\phi_q\cos(q_0x)$, where
the spatial average of $\phi(x)$ for the symmetric case is $\phi_0=\langle\phi\rangle=0$, and $\phi_q$ is its
modulation amplitude, while $\rho(x)$ is taken without any spatial modulations and is equal to its spatial average, $\rho_0=\langle\rho\rangle$~\cite{Garel82,Ohta_Kawasaki86,Villain_98,Villain_Netz_98}.
Taking a variation with respect $\phi_q$, it can
be shown that for the most dominant mode,
$q=q_0$, its amplitude $\phi_q$ satisfies:
\begin{equation}
\phi^2_q=4\rho_0^3\bigg(\tau-\frac{1}{\rho_0}\bigg) \, .
\label{phi^2_q}
\end{equation}
From eq~\ref{phi^2_q}, we get a condition for the extent
of the lamellar phase by requiring that $\phi_q^2\ge 0$, and
the critical temperature is obtained at $\phi_q=0$:
\begin{equation}
\tau_c=1/\rho_0 \, .
\end{equation}
The above equation agrees with $\tau_\textrm{c}=1$ for $\rho_0=1$. Furthermore, for
$\rho_0\rightarrow0$ we get $\tau_\textrm{c}\rightarrow\infty$ implying that the only possible phase
at low solute concentrations is the disordered phase.

\subsection{Induced Period Doubling in $\rho$}

\begin{figure}[h!]
  \includegraphics[width=0.45\textwidth]{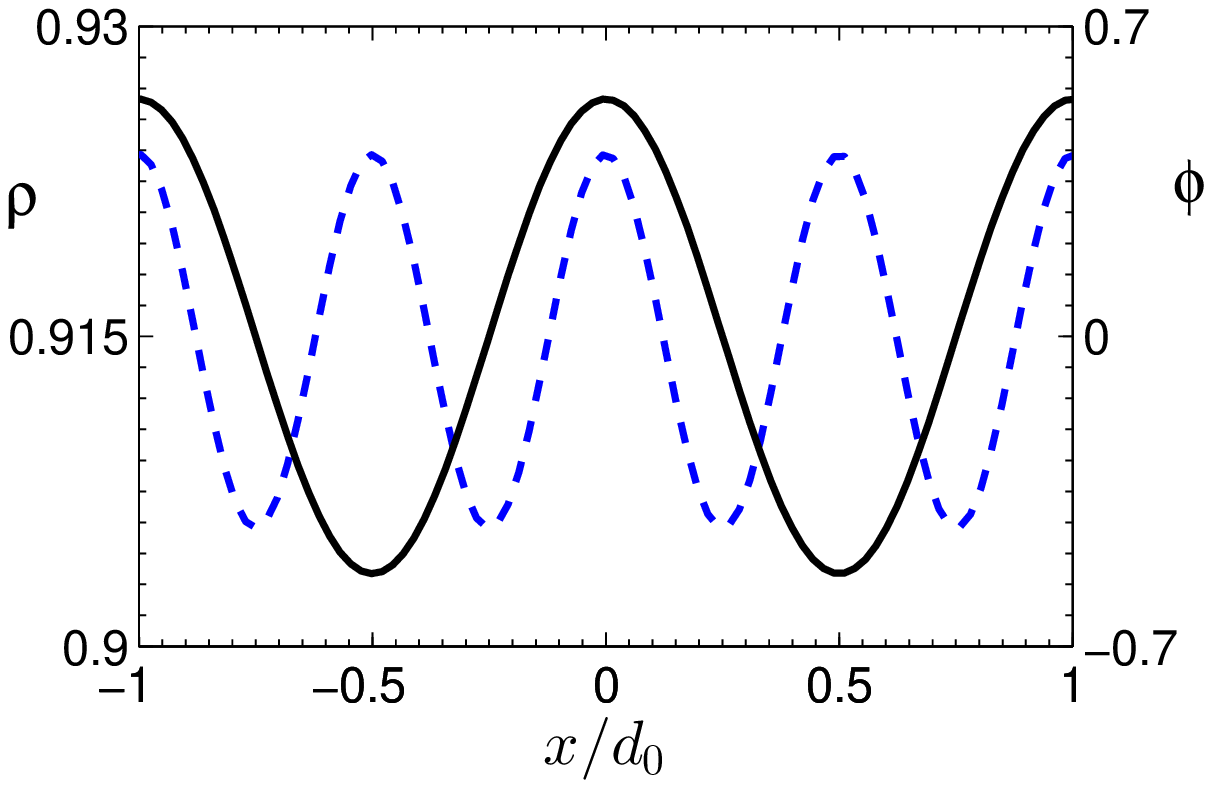}   
  \caption{\textsf{(colored online) The spatial modulation of $\rho$ (dashed line) and $\phi$ (solid line)
are plotted as function of $x/d_0$, with $d_0=2\pi/q_0$. The parameter values: $\tau=1.2$, $\chi=5.5$, $\mu_\phi=0$ and $\mu_\rho=0$
result in $\langle\rho\rangle=0.915$ and $\langle\phi\rangle=0$.
The oscillations in $\rho$ have half the wavelength of the
$\phi$ ones. The length scale of the $\rho$ modulations ($\rho_q=0.0091$) is much smaller
than the $\phi$ ones ($\phi_q=0.538$). The other parameter values are $H=K=1$ and $q_0=1/\sqrt{2}$.}}
\label{fig2}
\end{figure}

The modulations in $\phi$ cause the overall BCP density, $\rho$, to modulate as well.
Interestingly, the modulations in $\rho$ are
found to have half the wavelength of the $\phi$ modulations. We would like to explain how
this period doubling phenomenon is manifested in our formalism, and note that it has been already suggested and is not unique only to the lamellar phase~\cite{Helfand-Tagami71,Naugthon02}.

In figure~\ref{fig2} such
oscillations in $\phi$ and $\rho$ can be clearly seen, where  the amplitude of the $\rho$ oscillations is
much smaller than the $\phi$ ones.
The reason for such oscillations is the nonlinear coupling between the $\phi$ and $\rho$
order parameters. As we restrict ourselves to the case of symmetric interactions between A/S and B/S, $v_{\phi\rho}=0$
in eq~\ref{eq_full-free-energy},  this coupling originates in our model only from the entropy terms. Note that in figure 2 and in all subsequent figures we have chosen for convenience $H=K=1$ and $q_0=1/\sqrt{2}$.

\begin{figure}[!ht]
  \includegraphics[width=0.4\textwidth] {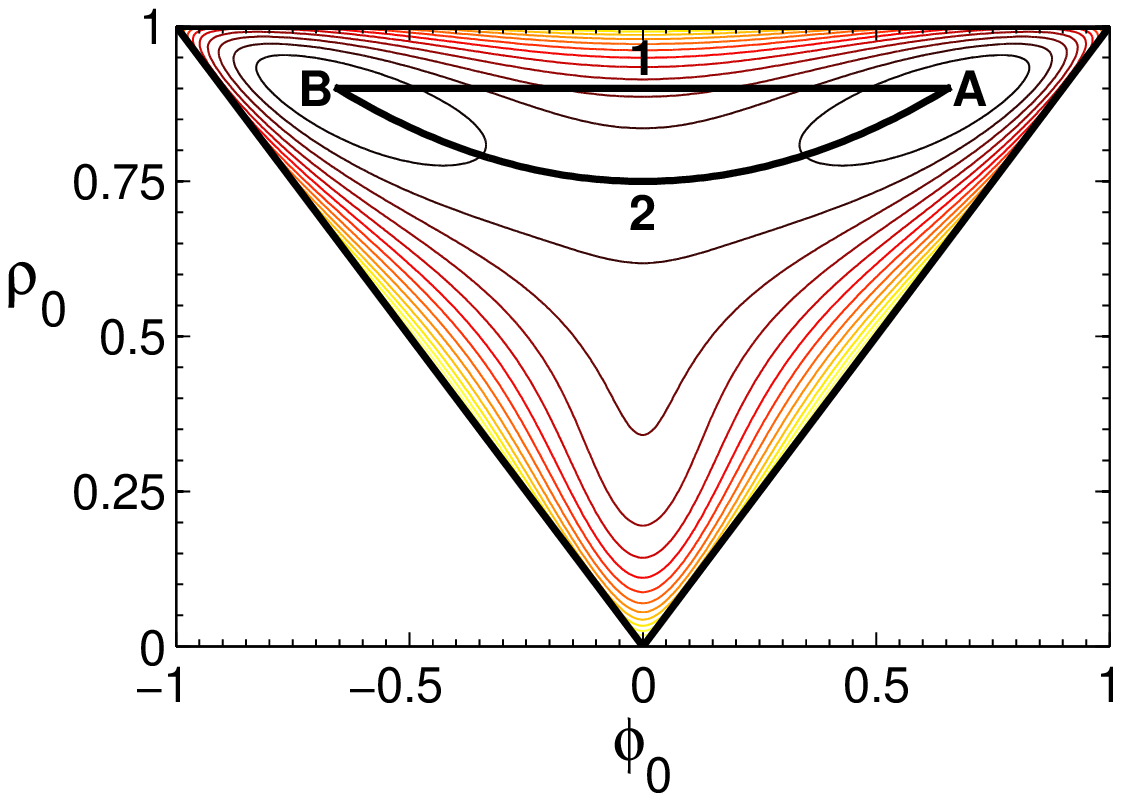}  
  \caption{\textsf{(colored online) Two possible routes from point $\textrm{A}$ to $\textrm{B}$,
plotted on the free-energy contour plot in the ($\phi_0$,$\rho_0$) plane, with
$\chi=4.2$, $\tau=1.45$, $\mu_\rho=0.05$ and $\mu_{\phi}=0$. `1' is a straight route in which $\rho$
does not change, while `2' is a curved one. The curved `2' path is energetically
preferred as it bypasses the energy barrier along the direct `1' path between A and B.}}
\label{fig3}
\end{figure}

In figure~\ref{fig3} we show the contour plot of the free energy for $\chi=4.2$ and
$\tau=1.45$. The spatial modulations in $\phi$ can be represented as an oscillatory
path between the two free-energy minima denoted as
$\textrm{A}$ and $\textrm{B}$. The path between the two points
will not take the direct route denoted by `1' for which $\rho=const$,
but rather a curved and longer route denoted by `2',
because the latter is energetically preferred as it avoids the energy barrier along the
direct path. Thus, oscillations in the $\phi$ order parameter induce oscillations in
$\rho$. The fact that the $\rho$ wavenumber is doubled ($2q_0$), as compared with the
$\phi$ wavenumber ($q_0$), can also be understood from figure~\ref{fig3}. For each cycle going from
$\textrm{A}$ to $\textrm{B}$ and back, $\phi$ completes one cycle, while $\rho$ completes two
cycles. We note that increasing $\chi$ will cause these oscillations to decrease their
magnitude as the potential well in the $\rho$ direction moves to higher values of $\rho$ and
becomes deeper.

We calculate the
induced $\rho$ modulations along one spatial direction, taken to be the $x$-direction; namely, $\rho=\rho(x)$ and $\phi=\phi(x)$.
An analytical (yet approximate) expression of the modulation amplitudes
can be obtained by expanding the free energy, eq~\ref{phi expansion}, in powers of
$\eta(x)=\rho(x)-\rho_{\rm D}$ around the value of the disordered phase, $\rho_{\rm D}$, to second order in $\rho$.
Because the
oscillations in $\phi$ are between two local minima, we can further expand eq~\ref{phi expansion}
to fourth order in $\phi$:
\begin{eqnarray}
\frac{g}{\kbt}&\simeq& -\frac{\tau}{2}\phi^2+\frac{\phi^2}{2\rho_{\rm D}}+\frac{\phi^4}{12\rho_{\rm D}^3}-
\frac{\phi^2}{2\rho_{\rm D}^2}\eta
\nonumber \\
& & +~\frac{1}{2}\bigg[\frac{1}{\rho_{\rm D}}+\frac{1}{1-\rho_{\rm D}}-\chi+\frac{\phi^2}{2\rho_{\rm D}^3}\bigg]\eta^2
\nonumber \\
&&+~K(\eta^{\prime})^2+\frac{H}{2}\bigg[\phi^{\prime\prime}+q_0^2\phi\bigg]^2 \, ,
\label{eq14}
\end{eqnarray}
where constant terms are omitted and $\rho_{\rm D}$ is given by:
\begin{equation}
\label{rho_D}
\frac{\chi}{2}(1-2\rho_{\rm D})+\ln\rho_{\rm D}-\ln(1-\rho_{\rm D})-\mu_{\rho}=0.
\end{equation}
Taking the variation of $g$ in eq~\ref{eq14} with respect to $\eta$ and using the lamellar single-mode
approximation, $\phi=\phi_q\cos(q_0x)$ for symmetric lamellae, $\langle \phi\rangle=0$, the inhomogeneous differential equation for $\eta(x)=\rho(x)-\rho_{\rm D}$ is:
\begin{align}
&2K\eta^{\prime\prime}-\frac{1}{2}\bigg[\frac{1}{\rho_{\rm D}}+\frac{1}{1-\rho_{\rm D}}-\chi+
\frac{\phi_q^2}{2\rho_{\rm D}^3}+\frac{\phi_q^2}{2\rho_{\rm D}^3}\cos(2q_0x)\bigg]{\eta}
\nonumber \\
&=-\frac{\phi_q^2}{4\rho_{\rm D}^2}\bigg[1+\cos(2q_0x)\bigg] \, ,
\label{e11}
\end{align}
where the linear term in $\eta$ originates from the quadratic term in the free energy.
This term has to be positive because we have made an expansion around the free-energy minimum. It means that the
homogenous  solution of $\eta(x)$ is a decaying function and is of no physical interest, since we are only looking for periodic solutions of the bulk phases. The inhomogeneous  solution $\eta(x)$, however, is caused
by  oscillatory cosine terms [RHS of eq~\ref{e11}].
To obtain an approximate solution we take $\rho_{\rm D}\simeq 1$ and
$\phi_q$ to be small, allowing us to neglect the sum of the two terms,
$\phi_q^2{(2\rho_{\rm D}^3)}^{-1}+\phi_q^2{(2\rho_{\rm D}^3)}^{-1}\cos(2q_0x)$, as compared with $(1-\rho_{\rm D})^{-1}$.
Then, eq~\ref{e11} simplifies to:
\begin{align}
&2K\eta^{\prime\prime}-\frac{1}{2}\bigg[\frac{1}{\rho_{\rm D}}+\frac{1}{1-\rho_{\rm D}}-\chi\bigg]{\eta}
\nonumber \\
&=-\frac{\phi_q^2}{4\rho_{\rm D}^2}\cos(2q_0x)-\frac{\phi_q^2}{4\rho_{\rm D}^2}\, .
\label{eta eqOFmotion}
\end{align}
Hence, the inhomogeneous solution has
the form $\eta=\eta_0+\eta_{2q}\cos(2q_0x)$, with~\cite{v2_not_0}:
\begin{align}
&\eta_0=\frac{\phi_q^2}{4\rho_{\rm D}^2\varepsilon}
\nonumber \\
&\eta_{2q}=\frac{\phi_q^2}{4\rho_{\rm D}^2(8Kq^2_0+\varepsilon)}\,\, ,
\label{eta_mod}
\end{align}
where $\varepsilon=\rho_{\rm D}^{-1}+{(1-\rho_{\rm D})}^{-1}-\chi$.
This solution represents the period doubling within our model.
It makes sense as the modulations in $\rho$ exist only when $\phi_q\neq0$, and may
appear at any value of $\rho_{\rm D}$. Larger densities or larger $K$ values
(high cost of the polymer/solvent interface) will make the modulations
smaller. The fact that $\eta_0=\langle\eta\rangle\ne0$ means that a
lamellar modulating phase will cause an
increase in the average solute density from the disorder phase value, $\rho_{\rm D}$.

\begin{figure}[!ht]
  \includegraphics[width=0.4\textwidth]{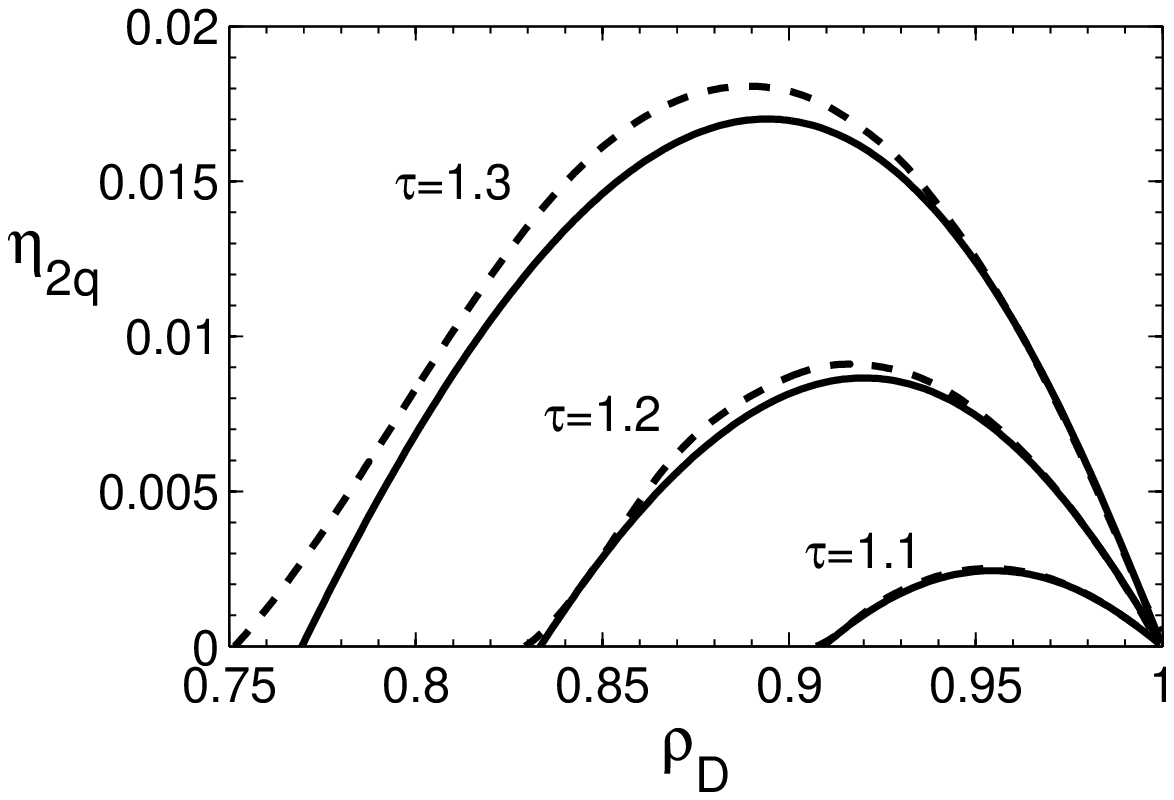}    
  \centering
  \caption{\textsf{Plot of $\eta_{2q}$ for symmetric lamellae as a function of $\rho_{\rm D}$ for three values of $\tau=1.1, 1.2, 1.3$, and with
  $\mu_\rho=\mu_\phi=0$.
The dashed line is the numerical solution of eq~\ref{eq_full-free-energy} and the solid
line is the analytical approximation of eq~\ref{eta_mod}.}}
\label{fig4}
\end{figure}

Comparison between the amplitude $\eta_{2q}$ as a function of
$\rho_{\rm D}$ as obtained by solving numerically eq~\ref{eq_full-free-energy},
and the approximated analytical one [using eqs~\ref{phi^2_q}, \ref{rho_D} and \ref{eta_mod}] is shown in figure~\ref{fig4} for various $\tau$ values, where symmetric lamellae with $\langle \phi_0\rangle=0$ are used. At small
$\tau$, the accuracy is very good since $\phi_q/\rho_{\rm D}$ is small. However, as $\tau$ grows so
does $\phi_q$, and the quality of the approximation deteriorates. Not surprisingly, as
$\rho_{\rm D}\rightarrow 1$, the modulations in $\rho$ disappear. For large $\tau$, the approximation worsens
not only when $\phi_q$ becomes large, but also next to the critical density $\rho_c =1/\tau$ when
$\rho_{\rm D}$ becomes smaller.

\section{Interfaces and Boundaries}

\subsection{Grain Boundaries of Solvent-diluted Lamellae} \label{wave doubeling interfaces}

\begin{figure}[!ht]
    \includegraphics[width=0.45\textwidth]{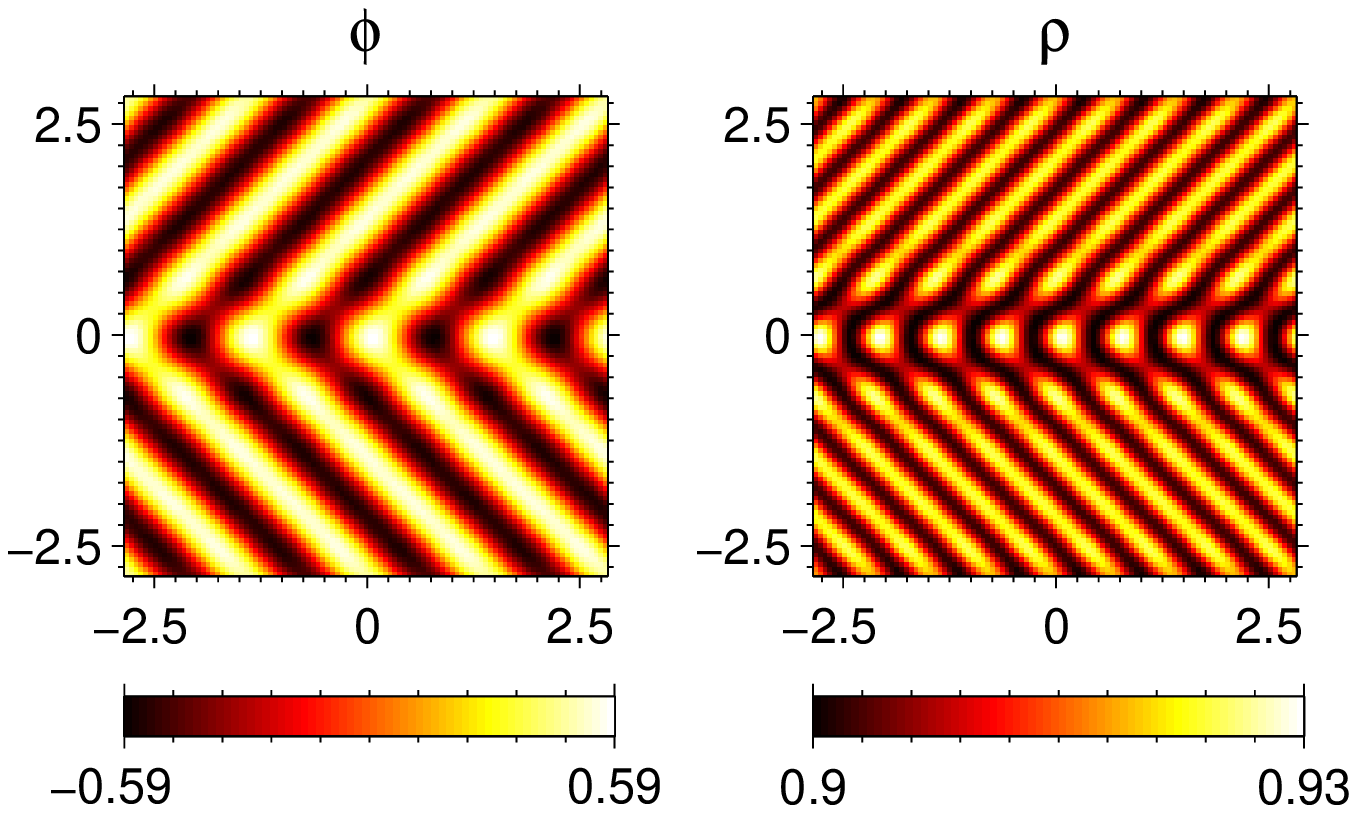} 
    \centering
    \linespread{1}
    \caption{\textsf{(colored online) Omega-shaped grain boundary, also called ``chevron", between two lamellar grains
that meet each other with a fixed angle. The angle is preset to be 90 degrees, and the parameter values are $\tau=1.2$,
$\mu_\rho=-0.004236$, $\chi=5.5$ and $\mu_\phi=0$. The axes are given in units of $d_0=2\pi/q_0$,
and the color bar accounts for variation of the $\phi$ (left) and $\rho$ (right)  order parameters.}}
    \label{fig5}
\end{figure}

\begin{figure}[!ht]

         \includegraphics[width=0.45\textwidth]{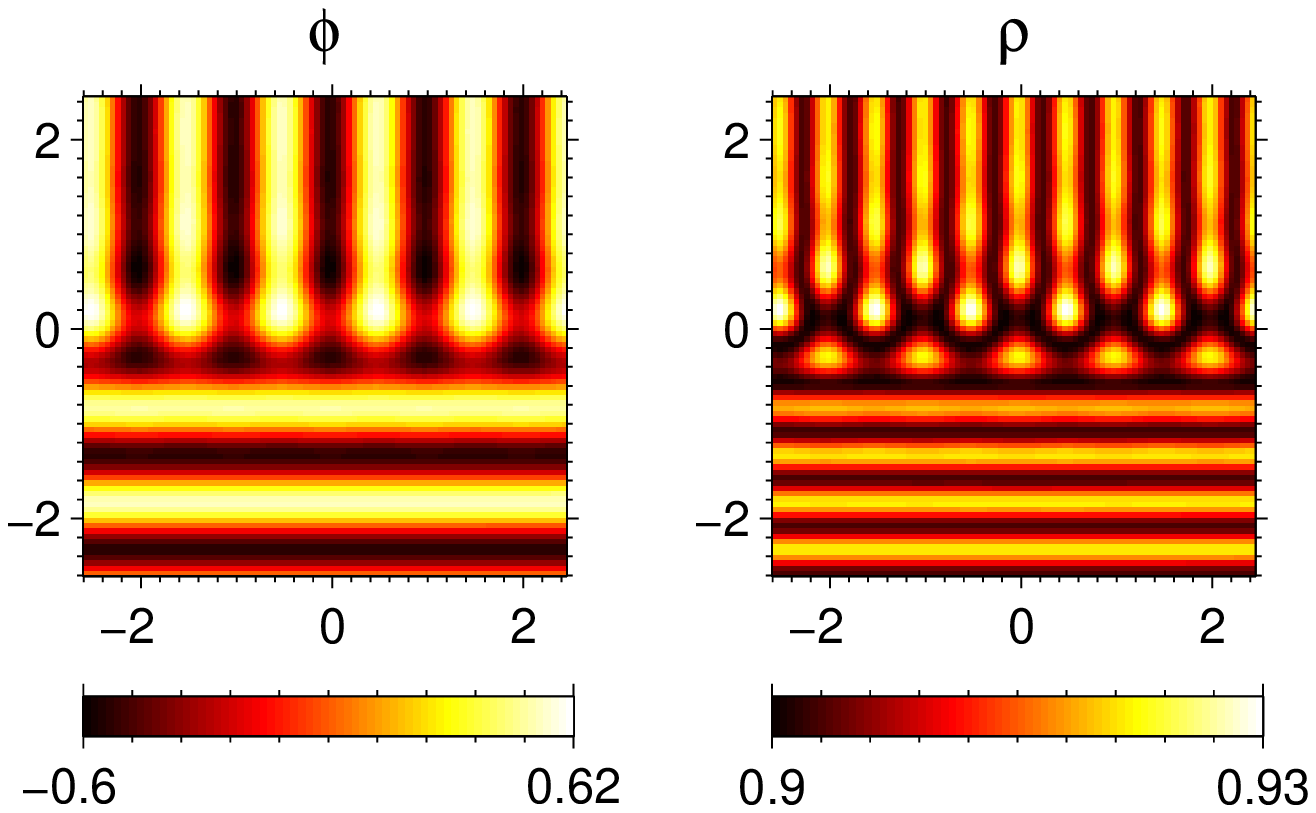} 
    \centering
    \linespread{1}
    \caption{\textsf{(colored online) T-junction grain boundary between two perpendicular lamellar domains.
    The parameter values are $\tau=1.2$, $\chi=5.5$, $\mu_\rho=-0.004236$ and
$\mu_\phi=0$. The axes are given in units of $d_0=2\pi/q_0$,
and the color bar accounts for variation of the $\phi$ (left) and $\rho$ (right)  order parameters.}}
    \label{fig6}
\end{figure}

We proceed by obtaining
grain boundaries separating lamellar domains of different orientations. This problem was considered in the past
for pure BCP boundaries~\cite{Netz_PRL_97,Tsori_chevron00} and here we extend it to solvent-diluted BCP lamellae.
The lamellar/lamellar grain boundaries
are obtained by minimizing numerically
the free energy in a simulation box, with boundary
conditions that are periodic on the vertical walls, while the top and bottom surfaces of the simulation box induce lamellar order with different orientations.
Such 2d patterns in $\phi$ and $\rho$ can be seen in figures
\ref{fig5} and \ref{fig6}. The left panels show the $\phi$ patterns, whereas the right panels
show the corresponding $\rho$ ones with a doubled wavenumber in their spatial periodicity, as was discussed in section~III.B.

Omega-shaped tilt
grain boundaries (figure~\ref{fig5}) and T-junction grain boundaries (figure~\ref{fig6})
are shown as examples of possible grain boundaries between lamellar phases and agree well with previous works on pure BCP systems~\cite{Netz_PRL_97,Tsori_chevron00}.
Due to the wavenumber
doubling effect one can see  in both figures ``bulbs" of higher density at the grain boundaries,
which is more pronounced for the $\rho$ patterns than for the $\phi$ ones. This result
is in agreement with experiments on dilute BCP systems~\cite{gido00} and calculations on blends of BCP and homopolymer acting as a solvent, which show that the homopolymer density is higher at the interfaces (e.g., T-Junction).
This probably can be explained in terms of a mechanism where the solvent molecules
accumulate at the location where the BCP
chains have large deformation and, hence, release some of  their strain at the interface~\cite{gido00}.

In general, the solvent is
enriched at the interface in order to dilute the unfavorable interactions between the two
polymer species. Since a BCP lamella contains two A/B interfaces, their distance is half the
lamellar spacing and can be obtained in other types of grain boundaries, as well as at interfaces
between two coexisting phases at equilibrium~\cite{Helfand-Tagami71,Naugthon02}.

\subsection{Substrate Chemical Patterning}
\label{chemically patterning}

Chemically and
topographically patterned surfaces with preferential local
wetting properties toward one of the two polymer blocks
result in a unique organization of BCP thin films and have been investigated thoroughly in experiments~\cite{Stoykovich05,Ruiz08,Bang09,Segalman01,Stein07,Li04,Voet11,Thebault12}. We will explore here the interplay between such chemical heterogeneities on surfaces and the structure
of BCP thin films.

\begin{figure}[!ht]
  \includegraphics[width=0.45\textwidth]{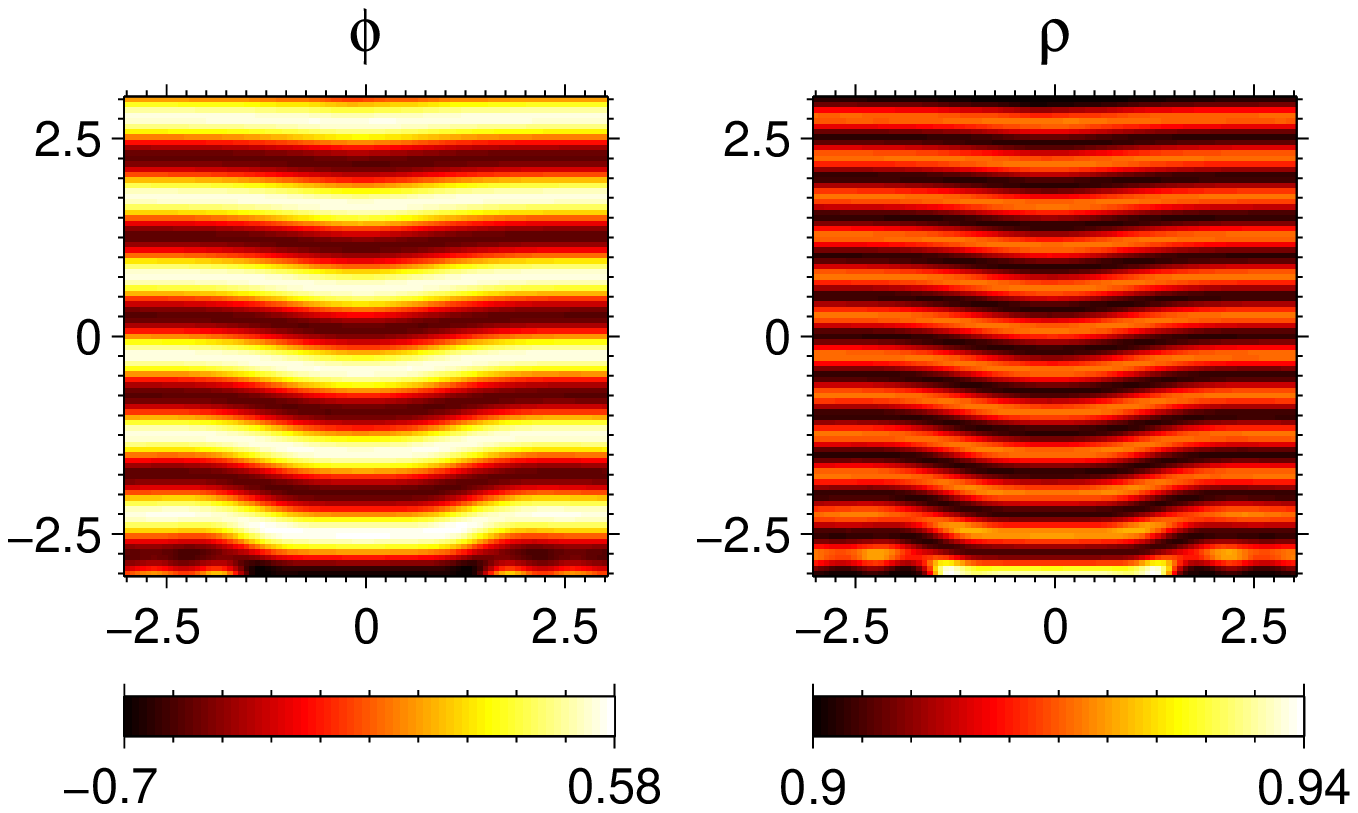}   
  \centering
   \linespread{1}
\caption{\textsf{(colored online) The  $\phi$ (left) and $\rho$ (right) order parameters
for a symmetric bulk lamellar phase, $\phi_0=0$, for a substrate that prefers
the B component in its central section as in eq~\ref{stripe}, while
we set periodic boundary conditions on the side walls.
The axes are given in units of $d_0=2\pi/q_0$,
and the color bar accounts for variation of each order parameter.
The patterns are produced using eq~\ref{eq_full-free-energy} with $\chi=5.5$, $\tau=1.2$
$\mu_{\rho}=-0.004236$, $\mu_{\phi}=0$, for which the lamellar phase the equilibrium one.
}}
  \label{fig7}
\end{figure}

The bulk BCP phase is placed in contact with a chemically patterned surface, modeled by two
surface interactions:  $\sigma_{\rho}$ and
$\sigma_{\phi}$~\cite{Tsori_01,para_perp}, which are coupled linearly to $\rho$ and $\phi$, respectively. They lead to a new
surface term $G_{\rm s}$ in the free energy
\begin{equation}
G_{\textrm{s}}=\kbt\int d^2\textbf{r}_s
\bigg[\sigma_\rho(\textbf{r}_s)\rho(\textbf{r}_s)+
\sigma_\phi(\textbf{r}_s)\phi(\textbf{r}_s)\bigg]\,,
\end{equation}
where $\textbf{r}_s$ is a vector on the 2d substrate. As an illustration of the chemical pattern influence on
a BCP lamellar phase, we examine the effect of the patterned substrate on structure and orientation of a parallel (L$_\parallel$)
lamellar phase as can be seen in
figure~\ref{fig7}. The substrate is constructed in such a way that its  field, $\sigma_\phi$, prefers
the $\textrm{B}$ component in the surface mid-section, while is neutral elsewhere:
\begin{equation}
    \sigma_\phi = \begin{cases}
                   0.5               & |x|\leq 1.5d_0\\
                   \\
                   0                 & |x| > 1.5d_0 \\
                \end{cases}
    \label{stripe}
\end{equation}
In addition, the field $\sigma_\rho$ is fixed to be zero on the entire substrate.
The upper bounding surface of the simulation box is taken as neutral, while the vertical walls obey periodic boundary conditions.
The deformation of the lamellar structure due to
the surface pattern are clearly seen close to the substrate and decays fast into the lamellar bulk as was previously studied for
pure BCP systems~\cite{Tsori_01}.

\subsection{The Free Interface}

\begin{figure}[!ht]
    \includegraphics[width=0.35\textwidth]{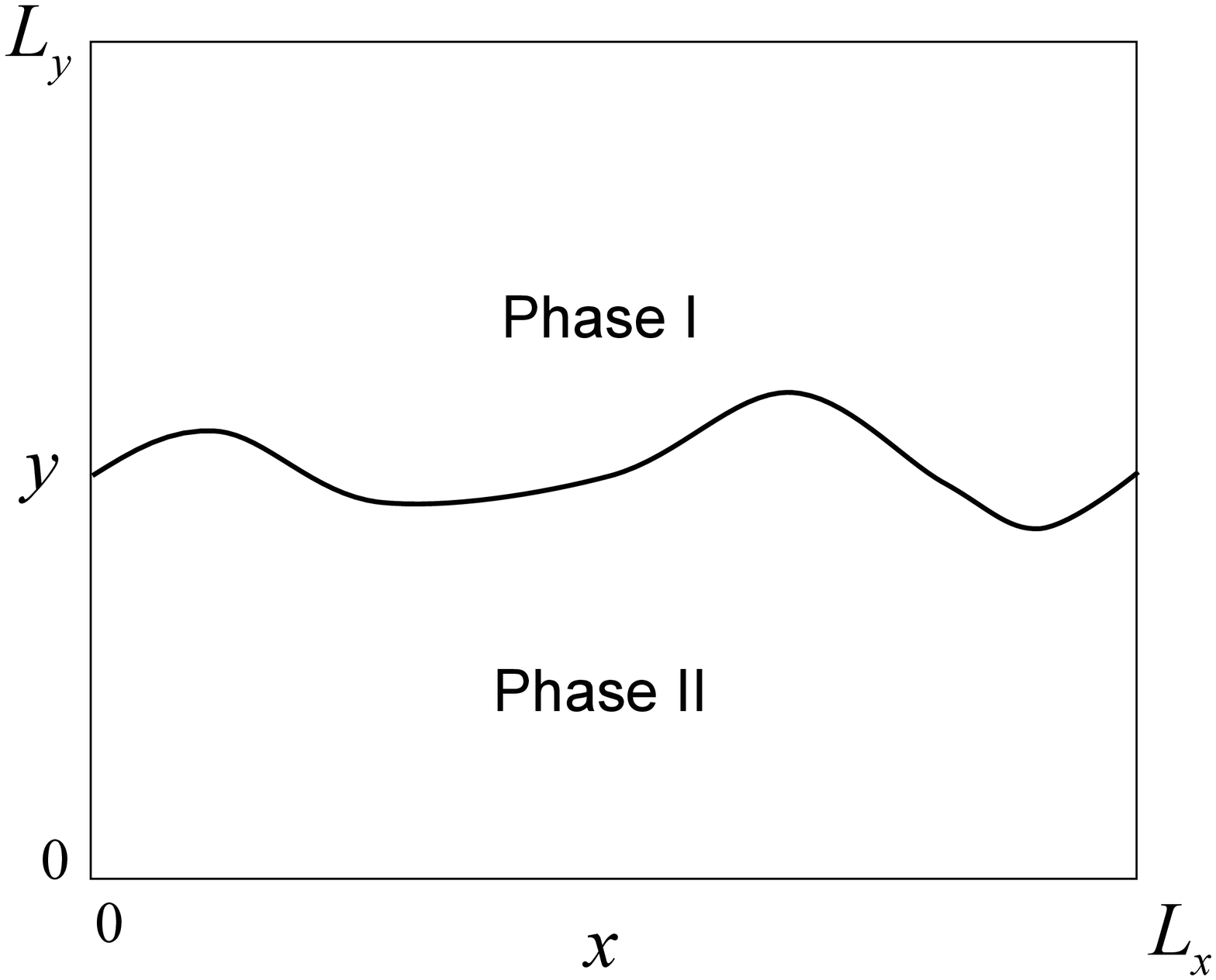} 
    \centering
    \linespread{1}
    \caption{\textsf{Schematic drawing of an  interface separating two coexisting phases, denoted as 'Phase I' and 'Phase II'.
    The 2d system volume is $V=L_x\times L_y$ and the interface projected area is $A=L_x$.}}
    \label{fig8}
\end{figure}

We can also address within our model and numerical scheme
the {\it free interface} between a thin  lamellar film in coexistence
with a bad solvent (vapor)~\cite{Matsen2012,X_Man_PRE12},
as is schematically plotted in fig~\ref{fig8}.
In particular, and motivated by experimental set-ups,
we explore deformations of such a free interface and their coupling with domain nucleations
as induced by different surface chemical patterings.

\begin{figure}[!ht]
    \includegraphics[width=0.45\textwidth]{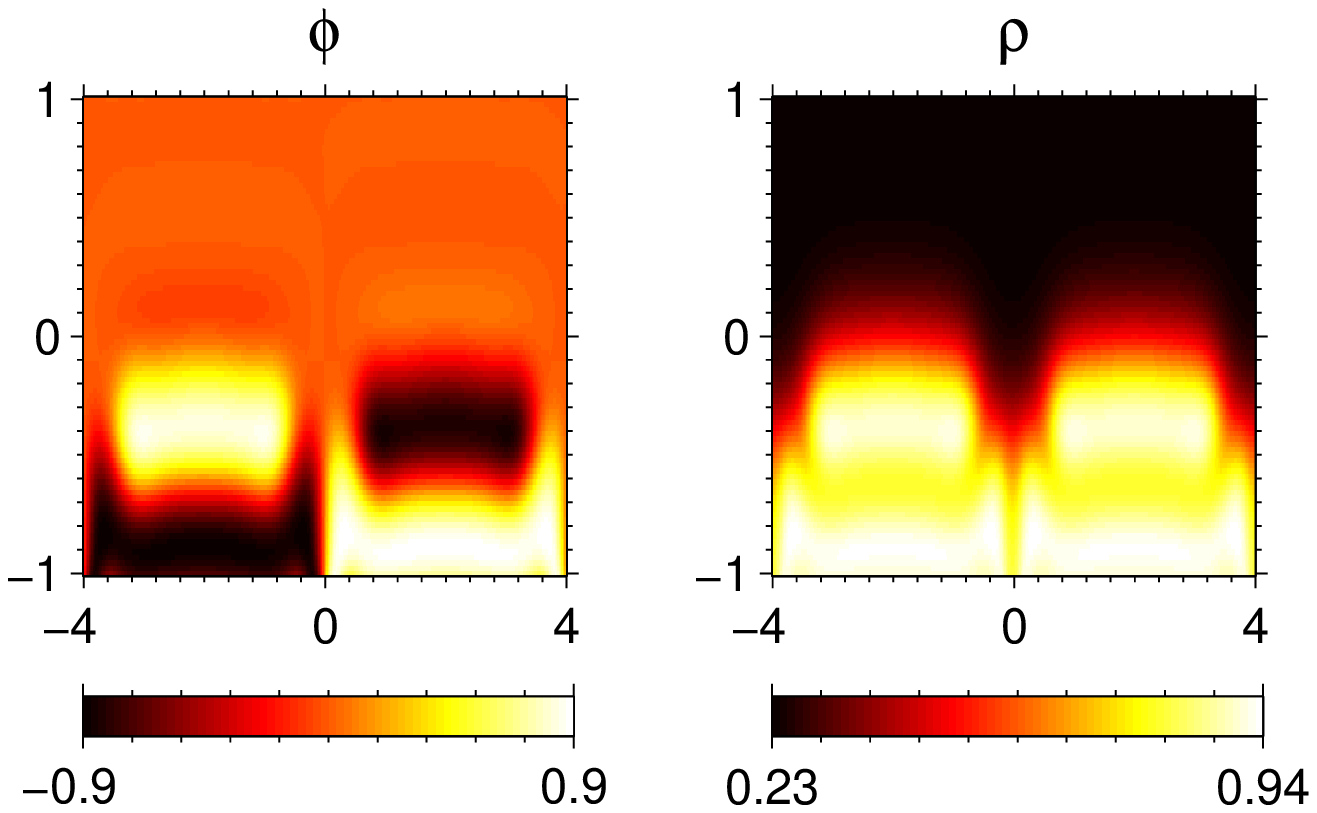}   
    \centering
     \linespread{1}
    \caption{\textsf{(colored online) Terrace formation both in $\phi$ (left) and $\rho$ (right)
    for a lamellar structure due to patterned substrate as in eq~\ref{eq22}.
 The system is at its solvent-lamellar coexistence with $\tau=1.8$, $\chi=4.1$, $\mu_\rho=-0.101682$ and $\mu_\phi=0$.
The axes are given in units of $d_0=2\pi/q_0$,
and the color bar accounts for variation of each order parameter.}}
    \label{fig9}
\end{figure}

In order to
allow deformations of the free interface, we consider specifically the case where the segregation in
$\rho$ is weak but that of $\phi$ is still strong. This is done by choosing $\tau=1.8$,
$\chi=4.1$ and $\mu_{\rho}=-0.101682$ (and $\mu_\phi=0$ for symmetric lamellae),
while recalling that the critical point values are
$\tau_c=1$ and $\chi_c=4$ (see section II). The effect of
surface-induced modulation on a weakly-segregated interface is seen in
figure~\ref{fig9}, where the lower surface is chosen to have the following step pattern:
\begin{equation}
    \sigma_\phi = \begin{cases}
                   -0.1               & \quad 0\leq {x}< 4d_0\\
                   \\
                   \quad 0.1               & -4d_0\leq {x}< 0\\
                \end{cases}
\label{eq22}
\end{equation}
and $\sigma_\rho=0$ for the entire substrate.
This patterning causes a defect formation at  the mid-point $x=0$, where the change in
$\sigma_\phi$ occurs. The defect, in turn, induces a deformation of the solvent/polymer (free) interface,
forming several terraces. The jumps in terrace height is about half of the lamellar
periodicity, in agreement with previous results obtained on free interfaces in contact with chemical patterns
using self-consistent field theory (SCFT)~\cite{X_Man_PRE12}.

\begin{figure}[!ht]
\includegraphics[width=0.4\textwidth]{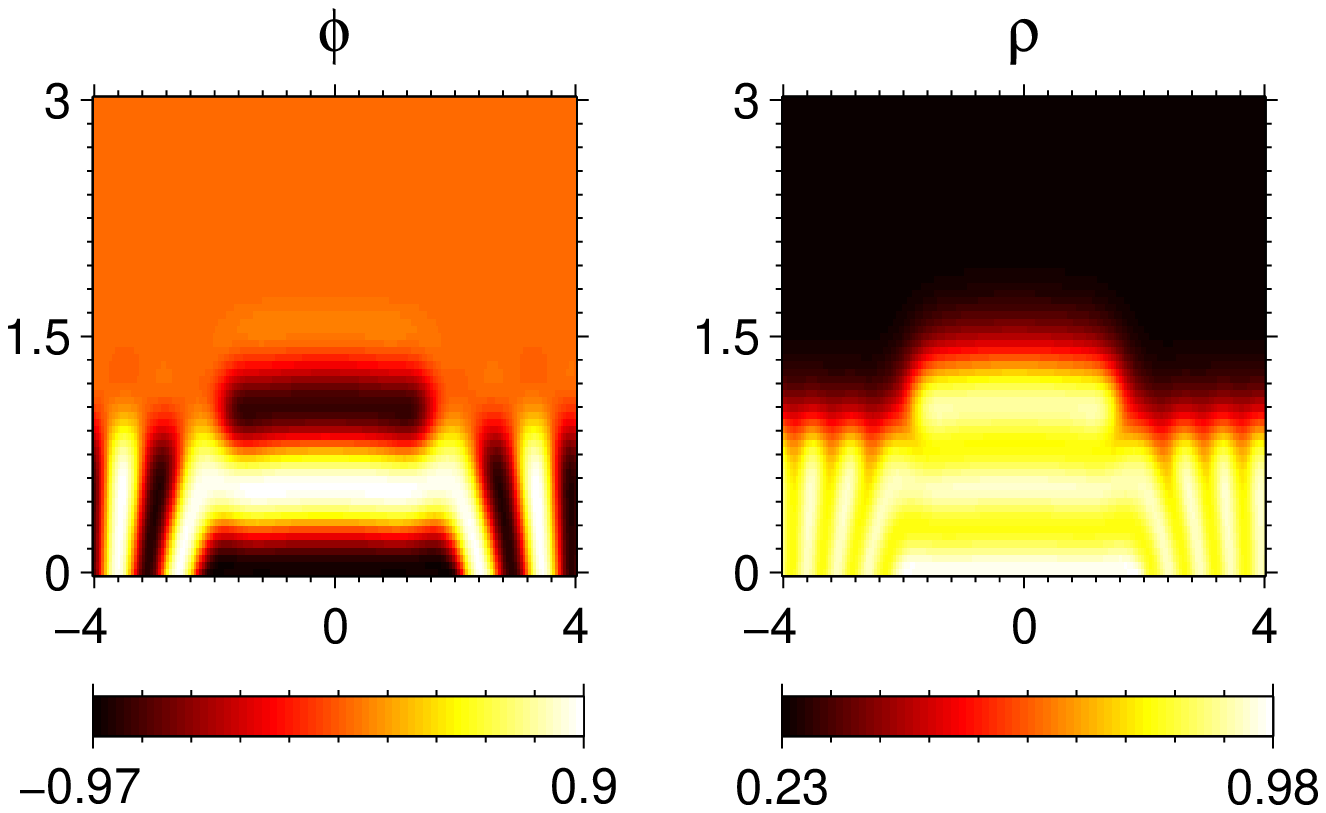}   
  \centering
   \linespread{1}
  \caption{\textsf{(colored online) Parallel (L$_\parallel$)  and perpendicular (L$_\perp$)  domains in
equilibrium with a pure solvent phase
($\sigma_{\rho}=0$) in presence of a surface pattern as in eq~\ref{eq23}. Terrace formation and deformation of
the free (solvent/BCP film) interface can be seen.
The system parameters are: $\tau=1.8$, $\chi=4.1$, $\mu_{\rho}=-0.101682$ and $\mu_\phi=0$.
The axes are given in units of $d_0=2\pi/q_0$,
and the color bar accounts for variation of the $\phi$ (left) and $\rho$ (right) order parameters.}}
  \label{fig10}
\end{figure}

A defect can also be obtained between the two lamellar orientations: parallel to the substrate ($\textrm{L}_{\parallel}$) and perpendicular one ($\textrm{L}_{\perp}$)
by  choosing a patterned substrate that
prefers the $\textrm{L}_{\parallel}$ one in its mid-section and an L$_\perp$ elsewhere:
\begin{equation}
    \sigma_\phi = \begin{cases}
                   0.1               & \quad\quad |{x}|\leq 2d_0\\
                   \\
                   0               & 4d_0> |{x}| > 2d_0\\
                \end{cases}
\label{eq23}
\end{equation}
and $\sigma_\rho=0$ for the entire substrate. As can be seen in figure~\ref{fig10}, this surface pattern causes a deformation of the free interface between the BCP (lamellar) and solvent phases. In the mid-section, L$_\parallel$ has a thickness of three layers (${y}\approx 1.5d_0$), and a terrace then separates the mid-section L$_\parallel$  from the side ones where the L$_\perp$ is the preferred orientation.
Moreover, the $\textrm{L}_{\perp}$ phase is deformed and tilted at the boundary with the L$_\parallel$ phase, and the width of this boundary increases as $\tau$ approaches its critical value.

\begin{figure}[!ht]
\includegraphics[width=0.45\textwidth] {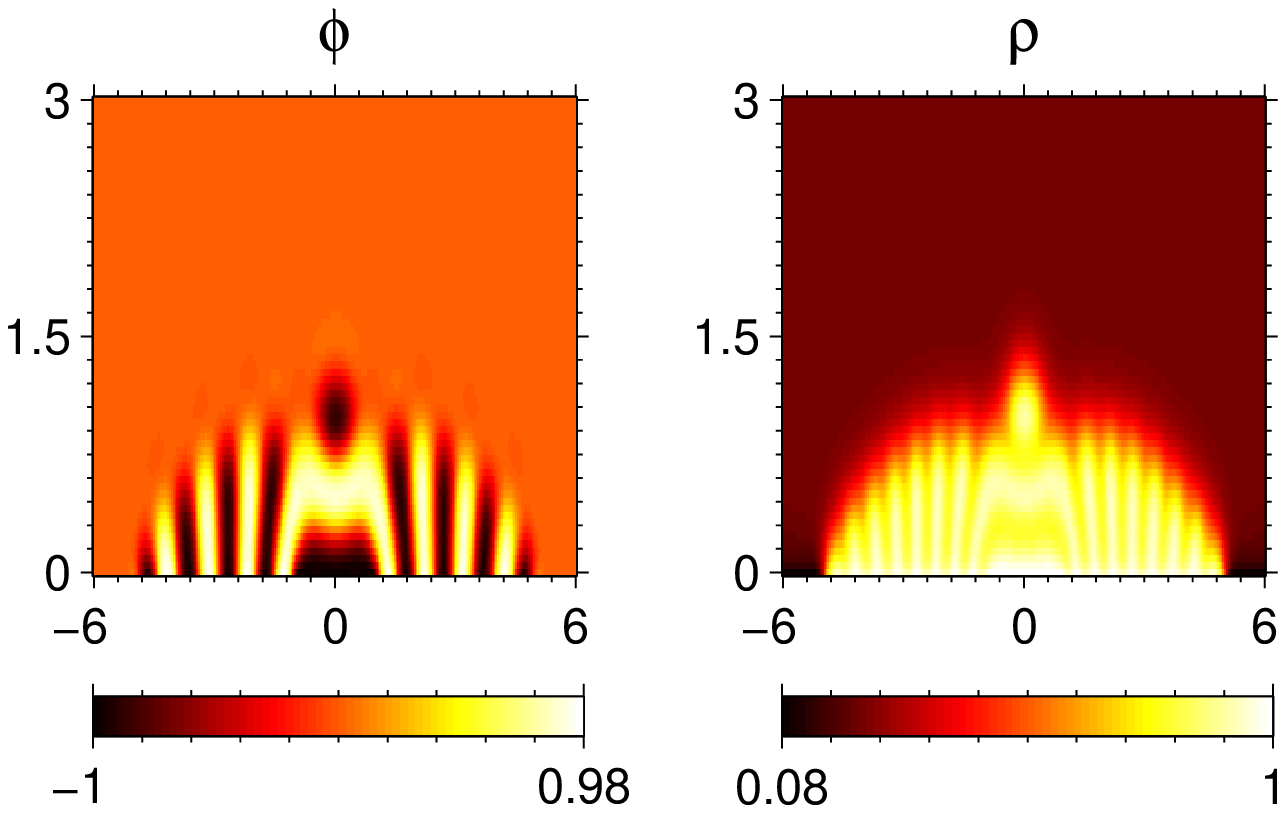}   
  \centering
   \linespread{1}
  \caption{\textsf{(colored online) A BCP lamellar droplet wetting a substrate. The $\phi$ (left) and $\rho$ (right)
   order parameters are plotted.
The wetting is formed using  a surface field that attracts the BCP to its mid-section
and repels it elsewhere, as in eq~\ref{eq24}.
Domains of different orientations are further induced inside the lamellar droplet by a $\sigma_\phi$
surface field [eq~\ref{eq25}]. The lamellar/solvent phase coexistence occurs for $\tau=1.8$,
$\chi=4.1$, $\mu_\phi=0$ and $\mu_\rho=-0.101682$.
The axes are given in units of $d_0=2\pi/q_0$,
and the color bar accounts for variation of each order parameter.}}
  \label{fig11}
\end{figure}

Another way of deforming the free interface is to induce a BCP droplet
by a patterned substrate in coexistence with the solvent phase.
For that purpose the substrate is separated into a central section that prefers the BCP with
$\sigma_\rho<0$ (BCP wetting condition), while on the rest of the surface $\sigma_\rho>0$, which
repels the lamellar phase (non-wetting condition that prefers the solvent).
In addition, by manipulation the $\sigma_\phi$ field,
we can induce different domains inside the same droplet. Such a case in which both $\textrm{L}_{\perp}$ and
$\textrm{L}_{\parallel}$ domains co-exist within the same BCP droplet is shown in figure~\ref{fig11},
with $\sigma_\rho$ and $\sigma_\phi$ given by:
\begin{equation}
    \sigma_\rho = \begin{cases}
                   -0.5               & \quad\quad|{x}|\leq 5d_0\\
                   \nonumber\\
                   \quad 0.5               & 6d_0 >|{x}| > 5d_0\\
                \end{cases}
\label{eq24}
\end{equation}
\begin{equation}
    \sigma_\phi = \begin{cases}
                   0.5               & \quad\quad|{x}|\leq d_0\\
                   \\
                    0               & 6d_0 > |{x}| > d_0\\
                \end{cases}
                \label{eq25}
\end{equation}

The patterning leads to central domain of the droplet in the $\textrm{L}_{\parallel}$ orientation surrounded by
two deformed $\textrm{L}_{\perp}$ domains that compensate between the height of the
$\textrm{L}_{\parallel}$ phase and the edges of the droplet. The morphologies and free-interface profiles found in this section result from substrate patterning, and are in agreement with
the ones obtained recently using a more
computationally intensive method of SCFT~\cite{X_Man_PRE12}.

\subsection{Surface Tension of Lamellar Phases}

As seen in the preceding sections, our model describes different types of coexisting phases and their interfaces. We proceed by analyzing the surface tension between a solute-rich lamellar phase
and a solvent-rich disordered
phase as function of the phase separation controlling parameters:  $\tau$ and $\chi$. An  illustration of the two coexisting phases in shown schematically in figure~\ref{fig8}, and our calculations are performed
for $2d$ systems of volume $V=L_x\times  L_y$. The surface tension between any two coexisting phases (I and II), is defined as $\gamma=[G^{\textrm{I},\textrm{II}}(V,A)-\frac{V}{2}(g^{\textrm{I}}_\textrm{b}+
g^{\textrm{II}}_\textrm{b})]/A$,
where $A=L_x$ is the projected area of the interfacial layer, and for convenience, each of the two phases occupies half of the total volume.

Coexistence between the two phases is achieved by tuning the appropriate chemical potentials.
The  free energy $G^\textrm{I}$ of a lamellar BCP phase
of volume $V/2$ is calculated numerically by imposing periodic boundary conditions on the side boundaries
and free boundary conditions on the upper and lower surfaces. The distance between top and
bottom surfaces of the box, $L_y$, is adjusted so that the free energy $G^\textrm{I}$ is minimized.
This occurs when $L_y$ is an integer multiple of the modulated periodicity.
Therefore, $G^\textrm{I}$ corresponds to a lamellar phase with no
defects, and with a free energy density
$g^\textrm{I}_\textrm{b}=G^\textrm{I}/(V/2)$. The same (but much simpler)
procedure is used to calculate the bulk free energy of the second (disordered) phase,
$g^{\textrm{II}}_\textrm{b}=G^{\textrm{II}}/(V/2)$. This procedure is
repeated iteratively until we find the chemical potential yielding two coexisting phases with
$g^{\textrm{I}}_\textrm{b}=g^{\textrm{II}}_\textrm{b}$.
When the simulation box $V$ is big
enough and for the proper chemical potentials, an initial guess of
the upper (I) and  lower (II) phases will converge into two coexisting phases
with a interface in between them.

\begin{figure}[!ht]
\includegraphics[width=0.4\textwidth]{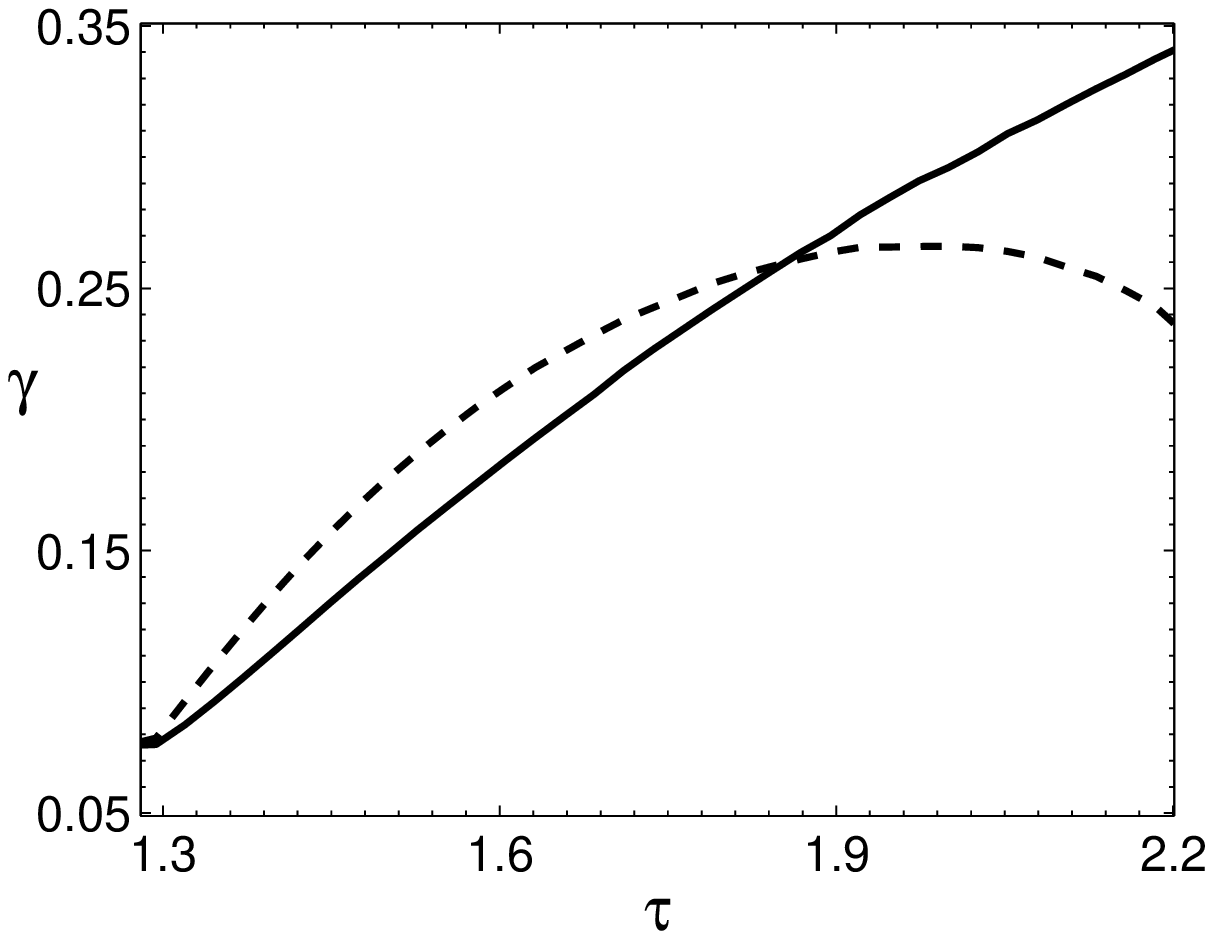}    
    \centering
    \caption{\textsf{Surface tensions, $\gamma_\parallel$ and $\gamma_\perp$, for the Lamellar-Solvent interface for two lamellar orientations, parallel (L$_\parallel$, dashed line) and perpendicular (L$_\perp$, solid line),
    as function of reduced temperature, $\tau$, and for fixed $\chi=4.5$. The lamellar phase exists only inside the interval $\tau>\tau_c\simeq 1.29$. Up to a crossover $\tau^*\simeq 1.848$,
$\gamma_{\perp}$ (solid line) has a lower value than $\gamma_{\parallel}$ (dashed line), and the perpendicular lamellae are preferred. While for $\tau>\tau^*$, $\gamma_{\perp}>\gamma_{\parallel}$ and L$_\parallel$  is preferred.}
}
    \label{fig12}
\end{figure}

We would like to compare the surface tension between a symmetric lamellar phase ($\phi_0=0$) and a solvent phase
where the lamellae
meet the L-S interface at different angles, for $\tau>\tau_c$. By choosing proper initial conditions we consider two limiting lamellar
orientations: parallel
to the interface ($\textrm{L}_{\parallel}$) and perpendicular one ($\textrm{L}_{\perp}$). In principle, other angles $\theta$ can be chosen as was done in ref~\cite{Netz_PRL_97} but we only consider the two extreme cases of $\theta=0$ and $\theta=90^\circ$. The corresponding surface tensions, $\gamma_\parallel$ and $\gamma_\perp$, are then computed as function of $\tau$ for $\chi=4.5$ and plotted on figure~\ref{fig12}.
As $\tau$ increases above $\tau_{\rm c}$, the segregation between the A/B components becomes stronger, causing an  increases in the
density change across the $\textrm{S}$ - $\textrm{L}$ interface. This leads to an increase of the two surface tensions, $\gamma_\parallel$ and $\gamma_\perp$, to increase.

\begin{figure*}[!ht]
    \includegraphics[width=0.9\textwidth] {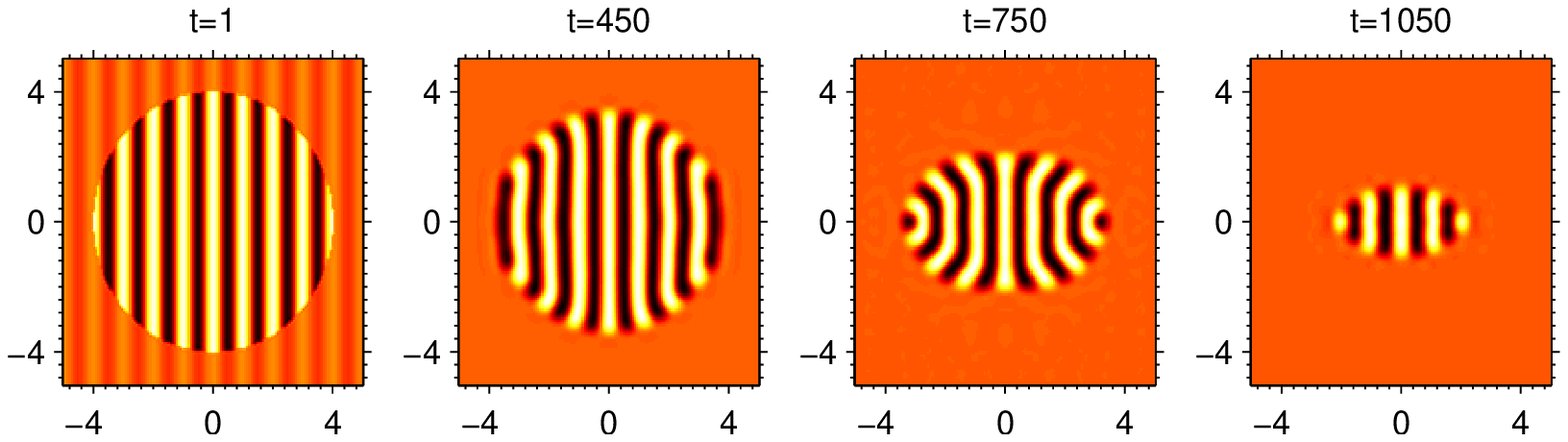}   
    \centering
    \caption{\textsf{(colored online) The temporal evolution of the $\phi$ order parameter of
    a lamellar BCP droplet in coexistence with a solvent
phase. From left to right: $t=1$, $450$, $750$ and $1050$, where the time step $t$  measures the number of numerical iterations. The parameters used are $\tau=1.6$, $\chi=4.5$, $\mu_\rho=-0.058629$, and $\mu_\phi=0$. As the
BCP droplet is metastable, its volume diminishes. The surface tension is anisotropic
because the lamellae meet the droplet interface with different angles.
It causes the original circular droplet to deform into a biconvex lens-shape, which is preferred energetically. The axes are given in units of $d_0=2\pi/q_0$.
}}
    \label{fig13}
\end{figure*}

It is worth noticing that $\gamma_\perp$ is a monotonically increasing function of $\tau$ while $\gamma_\parallel$ is non-monotonic.
However, while the segregation
between the $\textrm{A}$ and $\textrm{B}$ component grows,
the modulation amplitude reaches saturation, $\phi_q\rightarrow 1$, causing the width of the boundary
between A and B to diminish. This is in accord with ref~\cite{Netz_PRL_97} where similar trends with $\tau$ have been
reported at the lamellar-disorder interface (without a solvent).

Comparing the two orientations of the lamellar phase, it is seen that for low values of
$\tau$, the transverse orientation is
preferred at the interface, while for  larger $\tau$ values, $\textrm{L}_{\parallel}$ is the preferred one.
The crossover occurs at $\tau^*\simeq 1.848$ and indicates that the mere existence of the solvent interface may
induce a preferred direction of  bulk modulations. This result
implies that by incorporating steric repulsion (in which $\phi=\rho=0$ on any of the confining
surfaces),  the $\textrm{L}_{\parallel}$ phase is
preferred next to any neutral surface for large $\tau$ values.

\subsection{Shapes of Lamellar Droplets: Theory and Experiments}

Because  $\gamma_{\perp}$ and $\gamma_{\parallel}$  have very different dependence on $\tau$,
one might expect to see that an initial round
droplet containing a lamellar BCP becomes oval. For example, by choosing $\tau=1.6 <\tau^*$ and $\chi=4.5$,
$\gamma_{\perp}<\gamma_{\parallel}$ causes the $\textrm{L}_{\parallel}$ -- $\textrm{S}$ interface to
reduce its length, while the $\textrm{L}_{\perp}$ -- $\textrm{S}$ compensates and increases its
length. This is
seen in figure~\ref{fig13}. At initial times of the simulation [figure~\ref{fig13}(a)],
we have chosen a circular BCP droplet embedded in a solvent phase.
The droplet is circular and the lamellae are preset to orient vertically.
At progressive time steps of the simulation,
the lamellar orientation deforms so that the lamellae meet the interface at a right angle.
As in our simulations the BCP droplet is metastable, it diminishes in size but  not in a uniform way. Because the parallel boundary
diminishes faster ($\gamma_{\perp}<\gamma_{\parallel}$), the droplet undergoes a
continuous shape change, giving the droplet a biconvex lentil-shape at later time steps [figure~\ref{fig13}(d)].
If we continue the simulations even further, the droplet will eventually disappear, since the solvent phase is preferred.

We would like to compare our findings with experimental ones. Formation of lens-like BCP macro-domains embedded in a solvent matrix during solvent/polymer phase separation was observed experimentally~\cite{Hashimoto1994} for blends of poly(styrene-block-isoprene) and homopolystyrene acting as a bad solvent. The relationship between the macro-phase separation (solvent/BCP) and the self-assembly of the BCP inside the domains was reported and analyzed. Although we cannot offer a direct explanation of these experiments because our model is restricted to study lamellae in 2d, while in ref~\cite{Hashimoto1994} cylindrical phases of BCP are studied in 3d, the resemblance of our Fig. 15 with Fig. 5 of ref~\cite{Hashimoto1994} is striking, and may be regarded as 2d cuts through the 3d cylindrical BCP domain. We note that  in ref~\cite{Hashimoto1994} a similar explanation for the creation of lens-like  domains with orientation of the cylinders along the smaller lens axis was given, and is consistent with a difference in surface tension between the two orientations, $\gamma_\perp<\gamma_\parallel$.

More recently, addition of Au-based surfactant nanoparticles led to
change in shape and morphology for particles based
on poly(styrene-b-2-vinylpyridine) diblock copolymers \cite{Fredrickson2013}. The added nanoparticles are adsorbed at the interface between
block copolymer-containing droplets and the
surrounding amphiphilic surfactant. In turn, it causes a preferred perpendicular orientation of the BCP lamellae and led to distortion of the BCP droplets into ellipsoid-shaped ones. The system is more complex as it includes an additional component, but the explanation  presented by the authors for the distorted shape is
similar to ours and relies on the anisotropic surface tension as enhanced by the added Au nanoparticles.

In another study~\cite{Hashimoto1996}, lamellar domain formation has been reported following a
temperature quench from a disordered BCP
phase to the lamellar one. The formed lamellar domains are
lens-shaped with their axes along the smaller domain axis. It might be of interest to see if a
large change in the final temperature of quenching may cause the formation of lens domains with parallel-oriented lamellae, as is predicted by our study, where $\gamma_\perp$ and $\gamma_\parallel$ have different (and non-monotonous) temperature dependence. We also note that the nucleation of a droplet of stable cylinder phase from a metastable lamellar phase was examined
within the single-mode approximation for BCP melts in ref. \cite{Shi03}.

\section{Conclusions}

In this paper, we investigate bulk properties of solvent-diluted BCP phases, and a
variety of interfaces between coexisting BCP and solvent phases, restricting ourselves to phases with the 1d (symmetric lamellar) morphology. The main assumption made is
the dominant nature of a single $q_0$ mode close to the ODT, which allows us to write a simplified mean-field form of the free energy. Although the dominant $q_0$-mode
can be justified in the weak-segregation limit (close to ODT), we believe that many of the reported results are qualitatively correct also at stronger segregation.

To further simplify the model, the numerical investigations are conducted for the case where
the interactions between the A and B monomers of the BCP and the solvent (S) are identical. Namely, there is be no bilinear
enthalpic term as $\phi\rho$ in the free energy ($v_{\phi\rho}=0$) of eq~\ref{eq_full-free-energy}. Hence,
the coupling between the two order parameters, $\phi$ and $\rho$,
originates only from entropy of mixing that enhances solvent mediation
between the two incompatible species. This term is even in $\phi$ and its leading order is $\phi^2\rho$.

The coupling between $\rho$, the volume fraction of the solute, and $\phi$, the BCP relative A/B composition
leads to two interesting analytical results  valid in the low solvent limit, $\rho\approx 1$. First, the
critical point $\tau_c$  that determines the onset of BCP phases becomes $\rho$ dependent,
$\tau_c=1/\rho$.
Second, due to the non-linear nature of the coupling, modulations in $\phi$ induce modulations
in $\rho$, with a doubled wavenumber, as was explained in detail in section~III.B.
This phenomenon also results in formation of density bulbs at
the interface between a modulated BCP and a disordered phase
(figures~$\ref{fig5}$ and $\ref{fig6}$).

We show how
the presence of a chemically patterned substrate leads to deformations of the free interface
separating polymer-rich phase from a solvent-rich one.
The patterns can induce formation of terraces in lamellar BCP film and even formation of multi-domain droplets. Our results are in agreement with previous self-consistent field
theory calculations~\cite{X_Man_PRE12}.

It is of interest to remark on the rotational-symmetry breaking of BCP domains at the BCP/solvent coexistence.
We found that the surface tension of the parallel phase ($\gamma_\parallel$) is
higher than that of the transverse phase ($\gamma_\perp$) for values of $\tau$ close to $\tau_c$,
while the opposite occurs
for large $\tau$ values in agreement with ref~\cite{Netz_PRL_97}.
This crossover causes a BCP lamellar phase to change its
orientation relative to the interface as one changes $\tau$, and should be taken into
consideration, as it may be used to induce some preferred direction or interfere with such an
attempt. This difference in surface tension causes BCP droplets to deform. We believe that
it represents a general phenomenon that can be applied to other situations, such as the shape of domains
composed of a hexagonal
BCP phase coexisting  with a solvent phase.

{\bf Acknowledgements.~~~} {We thank H. Orland and X.-K. Man for many useful discussions. This work
was supported in part by the Israel Science Foundation under grant no. 438/12 and the US-Israel Binational Science Foundation
(BSF) under Grant No. 2012/060.}

\appendix

\section{Numerical Procedure }

The conjugate gradient (CG) method is a well-known numerical algorithm designed to find a local minimum of a smooth,
multi-dimensional nonlinear function \cite{NR}. In our case it was applied to minimize the Ginzburg-Landau free-energy expansion, given in eq~\ref{eq_full-free-energy}. The main reason for
using the CG method is in its convergence efficiency. For a parabolic minimization of a function that depends on $N$ variables,
the number of iterations can be reduced from $N^2$ to about linear in $N$.

We used a discrete $L_x\times L_y$ simulation box where the order parameters $\phi$ and $\rho$ depend on the discrete 2d lattice points $(x_i,y_j)$. The total energy is calculated as a sum over all sites, where the differences between the order parameter in one site and its neighboring sites is used to estimate the partial derivatives using their discrete form while penalty functions are used to avoid non-physical values of the order parameters.

Two types of boundary conditions are used in the numerical procedure. The first are periodic boundary conditions used to simulate bulk behavior. In the second case, the free energy is coupled to some surface field that can be uniform or represents a surface pattern.

As our model does not include random fluctuations,
the initial guess of the order parameters, $\phi$ and $\rho$, plays an important role.
In some cases a random initial guess is used to make the solver converge to the absolute minimum
(which is very time demanding), while in other cases a  well chosen initial guess is used
to speed convergence, especially when it is applied to model the interface between two coexisting phases. We
repeated the numerics by starting from several initial conditions in order to check that the convergence is toward the global free-energy minimum.

When the thermodynamics dictates that only one phase is at thermodynamical
equilibrium, the average value of the order parameters can be adjusted by
changing the chemical potential related to them. In a coexistence region of
two (or more) phases defined by setting the chemical potentials to their proper values,
the total BCP amount is not conserved for two-phase coexisting phases during the numerical iterations of the CG algorithm. However, the local convergence of the
lamellar phase and corresponding interface is much faster and our results indicate qualitatively the system state.

\newpage

%


\begin{thebibliography}{99}
\bibitem{v2_not_0}
{Note that if we include the asymmetric $v_{\phi\rho}$ term in eq~\ref{eq_full-free-energy}, this will add a 1st harmonic term in
the expansion of $\eta(x)$ of the form: $\eta_q\cos(q_0 x)$ with $\eta_q=2v_{\phi\rho}\phi_q/(4Kq_0^2+\varepsilon)$}
\end{thebibliography}
\end{document}